\documentclass[twocolumn,
			showpacs,
			preprintnumbers,
			amsmath,
			amssymb,
			aps,
			prl,
			floatfix,
			reprint,
			superscriptaddress]{revtex4-1}

\usepackage[ansinew]{inputenc}
	
\usepackage{graphicx}
\usepackage{dcolumn}
\usepackage{bm}
\usepackage{color}
\usepackage{verbatim}

\usepackage{textcomp}

\newcommand{\heidelberg}{Physikalisches Institut, Universit{\"a}t Heidelberg, Im Neuenheimer Feld 226, 69120 Heidelberg, Germany}
\newcommand{\ill}{Institut Laue-Langevin, BP~156, 6~rue Jules Horowitz, 38042 Grenoble Cedex~9, France}
\newcommand{\ati}{Atominstitut, Technische Universit{\"a}t Wien, Stadionallee~2, 1020 Wien, Austria}

\newcommand{\frm}{FRM II, Technische Universit{\"a}t M{\"u}nchen, Lichtenbergstra{\ss}e 1, 85748 Garching, Germany}
\newcommand{\thtuwien}{Institute for Theoretical Physics, Vienna University of Technology, Wiedner Hauptstra{\ss}e 8-10, 1040, Vienna, Austria}

\newcommand{\beq}{\begin{equation}}
\newcommand{\eeq}{\end{equation}}
\newcommand{\bea}{\begin{eqnarray}}
\newcommand{\eea}{\end{eqnarray}}

\usepackage{ulem}

\DeclareRobustCommand{\promille}{%
  \ifmmode
    \text{\textperthousand}%
  \else
    \textperthousand
  \fi}

\newcommand{\ket}[1]{\left\vert#1\right\rangle}

\begin{document}

\title{Gravity Resonance Spectroscopy Constrains Dark Energy and Dark Matter Scenarios}

\author{T.~Jenke} \email{tjenke@ati.ac.at} \affiliation{\ati}
\author{G.~Cronenberg} \affiliation{\ati}
\author{J.~Burgd{\"o}rfer} \affiliation{\thtuwien}
\author{L.A.~Chizhova} \affiliation{\thtuwien}
\author{P.~Geltenbort} \affiliation{\ill}
\author{A.N.~Ivanov} \affiliation{\ati}
\author{T.~Lauer} \affiliation{\frm}
\author{T.~Lins} \altaffiliation[Now at ]{\frm} \affiliation{\ati} 
\author{S.~Rotter} \affiliation{\thtuwien}
\author{H.~Saul} \altaffiliation[Now at ]{\frm} \affiliation{\ati} 
\author{U.~Schmidt} \affiliation{\heidelberg}
\author{H.~Abele} \email{abele@ati.ac.at} \affiliation{\ati}

%\author{H.\,Abele}
%\email{abele@ati.at}
\altaffiliation[Now at ]{\ati}
%\affiliation{\ati}

\date{\today}


\begin{abstract}
We report on precision resonance spectroscopy measurements of quantum states of ultracold neutrons confined above the surface of a horizontal mirror by the gravity potential of the Earth.
Resonant transitions between several of the lowest quantum states
are observed for the first time. These measurements demonstrate, that Newton's inverse square law of Gravity
is understood at micron distances on an energy scale of~$10^{-14}$~eV.
At this level of precision we are able to provide constraints on any possible gravity-like interaction.
In particular, a dark energy chameleon field is excluded for values of the coupling
constant~$\beta > 5.8\times10^8$ at~95\% confidence level~(C.L.), and an attractive (repulsive) dark matter axion-like spin-mass coupling is excluded for
the coupling strength $g_sg_p > 3.7\times10^{-16}$~($5.3\times10^{-16}$)~at a Yukawa length of~$\lambda = 20$~{\textmu}m~(95\% (C.L.).
\end{abstract}
\maketitle
Experiments that rely on frequency measurements can be performed with incredibly high precision.
One example is Rabi spectroscopy, a resonance spectroscopy technique to measure the energy eigenstates of quantum systems. It was originally developed by I.~Rabi to measure the magnetic moment of molecules~\cite{Rabi1939}. Today, resonance spectroscopy techniques are applied in various fields of science and medicine including nuclear magnetic resonance, masers, and atomic clocks.
%\\
These methods have opened up the field of low-energy particle physics with studies
of particle properties and their fundamental interactions and symmetries.
In an attempt to investigate gravity at short distances, we applied the concept of resonance spectroscopy to quantum states of very slow neutrons in the Earth's gravity potential~\cite{Jenke2011}.
Here, we present the first precision measurements of gravitational quantum states with this method that we refer to as gravity resonance spectroscopy (GRS).
The strength of GRS is that it does not rely on electromagnetic interactions. The use of neutrons as test particles bypasses the electromagnetic background induced by van der Waals and Casimir forces and other polarizability effects.

Within this work, we link these new measurements to dark matter and dark energy searches.
Observational cosmology has determined the dark matter and dark energy density parameters to an accuracy of two significant figures~\cite{PlanckCollaboration2013}.
While dark energy explains the accelerated expansion of the universe, dark matter is needed in order to describe the rotation curves of galaxies and the large-scale structure of the universe. The true nature of dark energy and the content of dark matter remain a mystery, however. The two most obvious candidates for dark energy are either Einstein's cosmological constant~\cite{Einstein1917} or quintessence theories~\cite{Wetterich1988,Ratra1988}, where the dynamic vacuum energy changes over time.
The resonant frequencies of our quantum states are intimately related to these models. If some as yet undiscovered dark matter or dark energy particles interact with neutrons, this should result in a measurable energy shift of the observed quantum states. One prominent dark matter candidate is the axion~\cite{Moody1984}, introducing a scalar-pseudoscalar coupling $g_sg_p$. Axion interactions in the so-called ``axion window'', given by the Yukawa length 0.2~{\textmu}m~$< \lambda <$~2~cm (corresponding to axion masses 10$^{-5}$~eV~$\leq m_a \leq$~1~eV), are still allowed by otherwise stringent constraints~\cite{Beringer2012}.
Recent reviews~\cite{Dubbers2011a} and~\cite{Tullney} cover this topic. The most restrictive limit on this product
$g_sg_p$ has been derived by combining the existing laboratory limit on the scalar coupling~$g_s$ with stellar energy-loss limits on the pseudoscalar coupling $g_p$~\cite{Raffelt2012}.

In this work, we determine experimental limits for a prominent quintessence theory, namely chameleon fields, and for the existence of axions at short distances. Other experiments have searched for spin-mass coupling at larger~$\lambda$ and had to extrapolate over several orders of magnitude.
\begin{figure*}[t]
	\centering
	\includegraphics[height=.17\textheight]{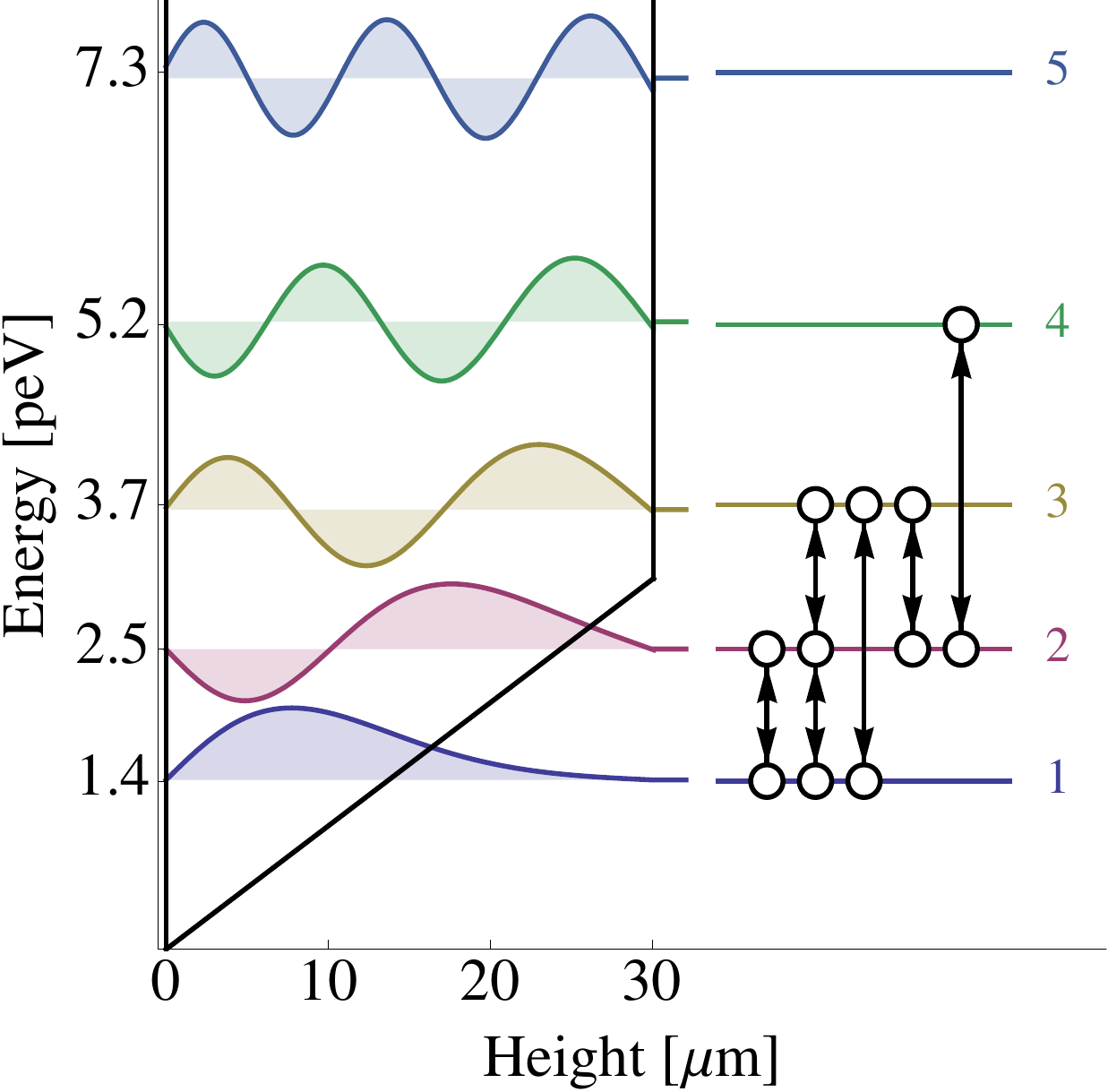}	
	\includegraphics[height=.17\textheight]{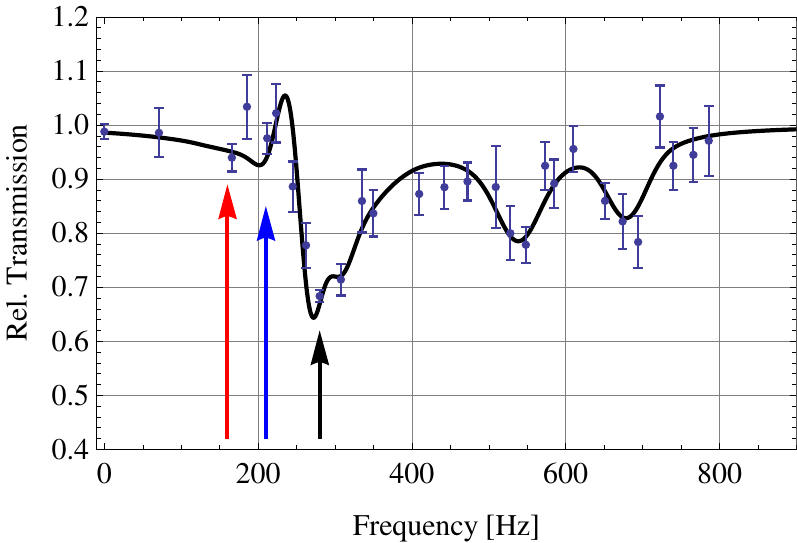}
	\includegraphics[height=.17\textheight]{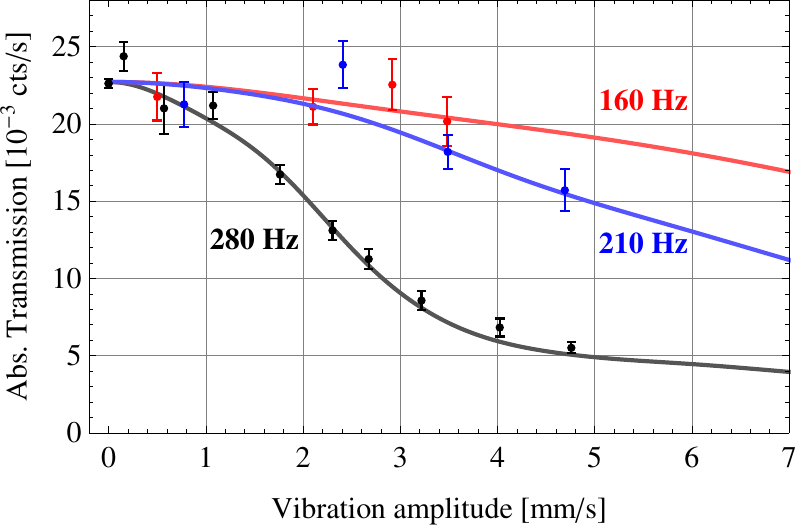}
\caption{Setup and results for the employed gravity resonance spectroscopy:
Left: The lowest eigenstates and eigenenergies with confining mirrors at bottom and top separated by 30.1~{\textmu}m.
The observed transitions are marked by arrows.
Center: The transmission curve determined from the neutron count rate behind the mirrors as a function of oscillation frequency shows dips corresponding to the transitions shown on the left.
Right: Upon resonance at 280~Hz the transmission decreases with the oscillation amplitude in contrast to the detuned 160~Hz. Because of the damping no revival occurs. All plotted errors correspond to a standard deviation around the statistical mean.
	\label{fig:GRS}}
\end{figure*}
In our experiment, we use very slow, so-called ultracold neutrons (UCN) confined in height direction~$z$ by two horizontal plates with separation~$l$.
The linear gravity potential leads to discrete, nonequidistant energy eigenstates as shown in Fig.~\ref{fig:GRS}, left, first measured in Refs.~\cite{Nesvizhevsky2002,Nesvizhevsky2005,Westphal2007}.
The eigenenergies~$E_k$ are based on the slit width~$l$, the neutron mass~$m_n$, the reduced Planck constant~$\hbar$, and the acceleration of the Earth~$g$.
The eigenfunctions - given by superpositions of Airy functions - additionally interact with a well-defined roughness of the upper horizontal plate.
This leads to an effective loss mechanism~\cite{Westphal2007} that results in state-dependent lifetimes~$\tau_k$, which decrease with increasing quantum number $k$ and, thus, provides a tool for state selection. As each transition can be addressed by its unique energy splitting, a combination of two states can be treated as a two-level system and resonance spectroscopy techniques can be applied.
In our case, we couple the two horizontal plates to a mechanical oscillator.
Alternatively, it has been proposed to drive transitions with alternating magnetic gradient fields~\cite{Baessler2011a}.
We apply sinusoidal mechanical oscillations with tunable frequency~$\nu$ and amplitude, and measure the corresponding neutron count rate.
This count rate drops close to the resonance condition $E_k - E_j = 2\pi\hbar \nu_{kj}$.
The Schr\"odinger equation that describes the UCN in the linear gravity potential between the two oscillating mirrors has the following Hamiltonian,
\begin{equation}
H = \frac{p^2}{2m_n} + m_ngz +
 V_F \Theta\left[ -z\! +\! f(t)
 \right] + V_F \Theta\left[z\! -\! l\! -\! f(t)
 \right].
\end{equation}
Here, $V_F$~corresponds to the Fermi pseudopotential of approximately~100~neV, which describes the interaction of UCN with the material of the walls~\cite{Ignatovich, *Golub91}. The Heaviside step function~$\Theta$ describes the boundary conditions, which oscillate with a frequency~$\nu$ and an amplitude~$A$: $f(t)=A\sin{\left(2\pi\nu t\right)}$.
The substitution $z \rightarrow z + f(t)$ transforms the time-dependent part of the Hamiltonian to
$W\left( z,t\right) = A \sin{\left(2\pi\nu t\right)} + i 2\pi\nu A \cos{\left(2\pi\nu t\right)} \frac{\partial}{\partial z}$~\cite{Abele2010}.
We use the ansatz
\beq
\psi \left(z,t\right) = \sum_k b_k\left(t\right) e^{-i\frac{E_k t}{\hbar}} e^{-i \phi_k} \ket{k}
\eeq
with time-dependent coefficients~$b_k(t)$ and phases~$\phi_k$. Here, $\ket{k}$~are the eigenstates of the undisturbed system with eigenenergies~$E_k$, see Fig.~\ref{fig:GRS}, left.
This leads to a system of coupled differential equations. To account for the state-dependent loss-mechanism, we add damping terms
$- \frac{1}{2\tau_k} b_k(t)$.
The restriction to a two-state system and a substitution into the rotating frame (where counter-rotating terms are neglected) leads to Rabi's differential equation with damping.
Since the transition frequencies~$\nu_{12}$ and~$\nu_{23}$ are not fully separated due to transit time broadening given by the inverse time-of-flight of the neutrons, these transitions have to be taken into account as a three-level cascade system~$\ket{1}\leftrightarrow\ket{2}\leftrightarrow\ket{3}$, for which an analytical solution exists. For a detailed derivation, see Ref.~\cite{Jenke2011a}.

In our experiment, we transmit UCN in between the two oscillating mirrors and measure the neutron flux behind the system as a function of the modulation frequency~$\nu$ and amplitude~$A$. On resonance $\left( \nu = \nu_{kj}\right)$, a so-called $\pi$ pulse induces transitions $\ket{k}\leftrightarrow\ket{j}$. Together with the asymmetric damping from the state-dependent loss mechanism, this leads to a change of the observed transmission.

We performed 135~transmission measurements for various modulation frequencies and amplitudes. This includes 17~measurements without oscillations and 19~measurements with polarized neutrons. The background rate of the detector~\cite{Jenke2013} is measured continuously between individual measurements and is found to be $\left(2.18 \pm 0.08\right)\times 10^{-3}$~s$^{-1}$.
The results for the neutron count rate behind the mirror system as a function of frequency are shown in Fig.~\ref{fig:GRS}, center: we identify the transition $\ket{1}\leftrightarrow\ket{3}$ at $\left(539.1^{+5.3\promille}_{-4.7\promille}\right)$~Hz and the so far unobserved $\ket{2}\leftrightarrow\ket{4}$ transition at $\left(679.5^{+2.0\%}_{-2.4\%}\right)$~Hz.
Furthermore, a deep drop of intensity is observed around $\nu = 280$~Hz, which is the result of the three-level cascade system $\ket{1}\leftrightarrow\ket{2}\leftrightarrow\ket{3}$ with transition frequencies $\nu_{12} = \left(258.2^{+7.3\promille}_{-8.6\promille}\right)$~Hz and $\nu_{23}=\left(280.4^{+1.2\%}_{-8.5\promille}\right)$~Hz, respectively. These two transitions overlap due to their finite width of 53~Hz, given by the inverse of the interaction time with the oscillating potential.
In order to display measurements obtained at the same oscillation frequency but slightly different oscillation amplitudes in the same figure, we normalize the measured count rate to the average rate without external oscillations. We then multiply the result by the quotient of the best fit value for the average oscillation amplitude of~$2.03$~mm/s and its actual value from the measurement. Additionally, an equidistant binning of size 20 Hz is used.
\\
To explore the dependence on the oscillation amplitude, we scan this parameter for three fixed frequencies 160, 210, and 280~Hz, corresponding to the arrows in Fig.~\ref{fig:GRS}, center. We find Rabi-oscillation curves, which are damped due to losses, as predicted theoretically (see Fig.~\ref{fig:GRS}, right). The theory curves shown are obtained from a fit to all~116 background-subtracted raw measurements with unpolarized neutrons.
This fit has ten free fit parameters. The three parameters $\nu_{12}$, $\nu_{23}$, and $\nu_{24}$ determine the transition frequencies. Energy conservation requires  $\nu_{12}$ + $\nu_{23}$ = $\nu_{13}$.
The three lifetimes~$\tau_2$, $\tau_3$, and $\tau_4$ account for the state-dependent loss mechanism, determined relative to the ground state.
The state populations~$c_2$, $c_3$~and~$c_4$ relative to state $|1\rangle$ define the initial conditions. Finally, an overall normalization parameter is used.
The $\chi^2_{min} = 114.7$ value found corresponds to a p-value~\cite{Beringer2012} of 26.6\%.
The transition frequencies are determined by the acceleration of the Earth and the slit height~$l$ between the two mirrors. Setting $g = 9.805$~m/s$^2$ to its local value, the $\chi^2$ still corresponds to a sufficiently good p-value of 16.7\%.

Our experiment was performed at the UCN installation PF2 at the Institut Laue-Langevin~(ILL). For a schematic diagram of the set-up, see Supplemental Material~\cite{supplemental}. The entire setup is mounted on a polished plane granite table, leveled to an absolute accuracy of better than $10$~{\textmu}rad and stabilized to a precision better than 1~{\textmu}rad. A {\textmu}-metal shield suppresses the coupling of residual magnetic field gradients to the neutron's magnetic moment. The whole experiment takes place in vacuum of approx.~10$^{-4}$~mbar.
At the entrance, a collimating system selects neutrons with a horizontal velocity of 5.7~m/s~$<v_x<$~9.5~m/s. The vertical boundary conditions are realized using a polished glass mirror at the bottom and a rough glass mirror at the top. The lower mirror has a roughness of less than 2~nm and a waviness of less than $\pm10$~nm; the upper one possesses a roughness of 3~{\textmu}m.
Mechanical spacers separate these mirrors by a distance of $l=27$~{\textmu}m.
The neutron mirror set-up is mounted on a piezo-based nanopositioning table with three degrees of freedom, the height~$z$, the tip, and the tilt angle. The table allows for vertical sinusoidal oscillations by applying a small sinusoidal voltage to the corresponding input.
For the spectroscopic measurements described in this Letter, a background-optimized counting detector with an efficiency of about 77\% is used~\cite{Jenke2013}. The mirrors have a length of~150 mm; the time of flight in this interaction zone determines the Rabi linewidth.
For spin-dependent searches, the modified detector has an entrance foil coated with soft iron.
The alignment of the polarization of the foil selects the spin direction of the neutrons.
The foil polarization has been measured separately to be 93\%.\\
\begin{figure}[t]
	\centering
	\includegraphics[width=.40\textwidth]{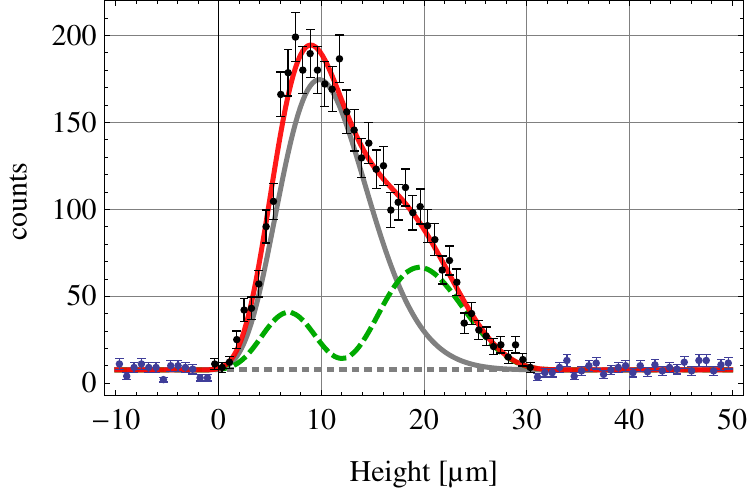}
	\caption{Spatial probability density of UCN in Earth's gravitational field times the number of counted neutrons  measured with a track detector~\cite{Jenke2013} behind the mirrors. A fit to the data (red line) gives the relative abundances of state~$|1\rangle$ (grey line) and $|2\rangle$ (green dashed line). No higher states are found.}
	\label{fig:cr39}
\end{figure}
To characterize our neutron mirror setup, we measure the probability density $\left|\psi\right|^2$ behind the system using track detectors with a spatial resolution of ca.~2~{\textmu}m~\cite{Jenke2013}. A fit to the data (see Fig.~\ref{fig:cr39}) gives a ground-state population of 70\%. No state population higher than $|2\rangle$ is observed, which validates the state selection process.

The sensitivities obtained for the energy measurements of all but the $\ket{2}\leftrightarrow\ket{4}$ transition are at the 10$^{-14}$eV level. The experimental error is dominated by the statistical uncertainty of the measurements. All known systematic effects lead to shifts well below this energy scale: the largest systematic influence arises from the surface disorder of the upper neutron mirror.
Because of the roughness, the slit width~$l$ cannot be measured accurately enough and is, therefore, treated as free parameter when expressing the resonance frequencies in terms of slit width and local value of~$g$. Extensive numerical calculations based on an explicit solution of the corresponding
scattering problem were carried out to validate this method at the present sensitivity level.
Systematic effects due to the limited control of the harmonic driving potential because of the eigenresonance at 122~Hz, the relative long-term frequency stability below 10$^{-5}$, and the inhomogeneity of the sinusoidal oscillation amplitude on the 10\%-level lead to uncertainties that are at least one order of magnitude smaller than our current statistical uncertainty.
Inclination changes of the setup were controlled on the {\textmu}rad level. The quality of the neutron mirrors regarding roughness and waviness lead to systematic effects below 10$^{-19}$eV.
We can safely ignore changes in the local acceleration of the Earth; tidal effects come in at the 10$^{-19}$eV level, and effects due to the Coriolis force are well below 10$^{-17}$eV.
In contrast to other neutral test particles such as atoms, neutrons possess an extremely small polarizability. Systematic effects due to van der Waals and Casimir forces are strongly suppressed to below the 10$^{-28}$eV level.
\\
With this remarkable level of control, the present experimental results allow us to search for any new kind of hypothetical gravitylike interaction at micron distances. At this natural length scale of the quantum states, the experiment is most sensitive (see Fig.~\ref{fig:GRS}, left).

\begin{figure}[t]
	\centering
	\includegraphics[width=.40\textwidth]{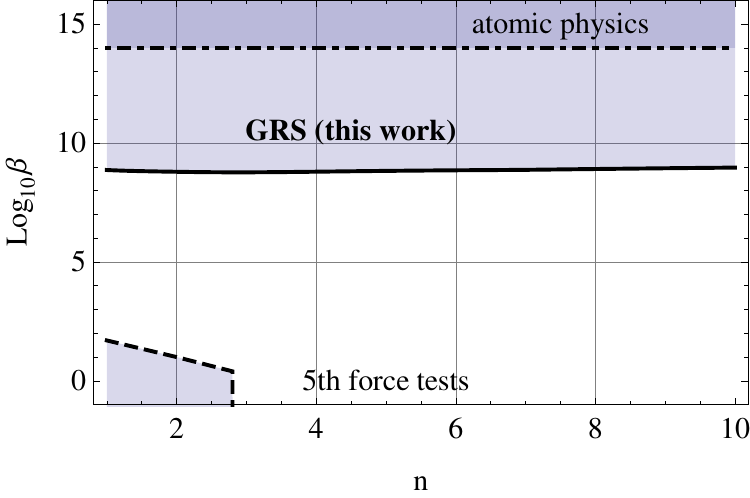}
	\caption{Exclusion plot for chameleon fields (95\% confidence level). Our newly derived limits (solid) are five orders of magnitude lower than the upper bound from precision tests of atomic spectra (dashed)~\cite{Brax2011}. Pendulum experiments~\cite{Adelberger2009} provide a lower bound (dotted).}
	\label{fig:chameleonexclusion}
\end{figure}

First, we address dark energy as a realization of quintessence theories with direct coupling to matter. A particularly appealing realization is the so-called chameleon scenario~\cite{Khoury2004a,Mota2006,Mota2007,Waterhouse2006}, where a combination of the potential $V(\Phi,n)$ of a scalar field~$\Phi$ and a coupling~$\beta$ to matter together with model parameter~$n$ leads to the existence of an effective potential $V_{\rm eff}$ for the scalar field quanta, which depends on the local
mass density~$\rho$ of the environment:
\beq
V_{\rm eff}=V(\Phi,n)+e^{\beta \Phi/M_{Pl}'}\rho.
\eeq
Here, $M_{Pl}'$ corresponds to the reduced Planck mass.
Our method directly tests the chameleon-matter interaction and does not rely on the existence of a chameleon-photon-interaction as other experiments do~\cite{Ahlers2008,*Gies2008,*Steffen2010,*Upadhye2012,*Rybka2010,*Wester2011}.
\\
The chameleon field potential for our setup is derived in Ref.~\cite{Ivanov2013}.
This result was obtained for the case of an ideal vacuum~($\rho = 0$), but remains valid at
room temperature and vacuum pressure of~10$^{-4}$~mbar.
We calculate bounds on the coupling constant~$\beta$ by comparing the transition frequencies with their theoretical values, which are proportional to the matrix elements $\nu_{kj}-\nu_{kj}^{\rm theo}\sim \beta(\langle k|\Phi|k\rangle-\langle j|\Phi|j\rangle).$
In the corresponding data analysis, the fit parameter for Earth's acceleration was fixed at the local value $g = 9.805$~m/s$^{2}$, while all other parameters were varied. The extracted confidence intervals for limits on the parameters $\beta$ and~$n$ are given in Fig.~\ref{fig:chameleonexclusion}.
The experiment is most sensitive at $2 \leq n \leq 4$ (visible only on a linear scale of Fig. 3), where a chameleon interaction is excluded for
$\beta > 5.8 \times 10^8$ (95\% C.L.).

The present limit is five orders of magnitude lower than the upper bound from precision tests of atomic spectra~\cite{Brax2011}. The parameter space is restricted from both sides, as other experiments provide a lower bound of $\beta < 10$ at $n<2$~\cite{Adelberger2009,Brax2011}.
In the future, an improvement of seven orders of magnitude would, thus, be necessary to exclude the full parameter space for chameleon fields for small~$n$.

\begin{figure}[t]
	\centering
	\includegraphics[width=.40\textwidth]{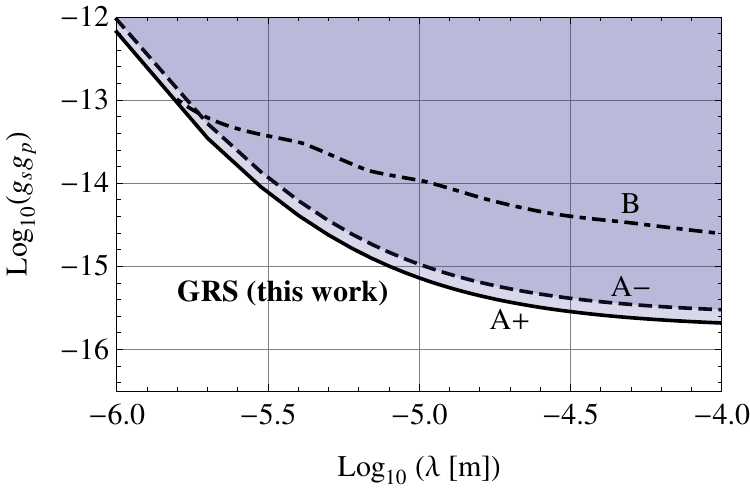}
	\caption{Limits on the pseudo-scalar coupling of axions (95\% confidence level). Our limit for a repulsive~(attractive) coupling is shown in a solid~(dashed) line marked with A+(A-). The limits are a factor of~30 more precise than the previous ones derived from a direct measurement at the micron length scale~\cite{Baessler2007} derived from our previous experiment with UCN marked B~\cite{Nesvizhevsky2005,Westphal2007}.}
	\label{fig:axionexclusion}
\end{figure}

Second, we perform a direct search for dark matter. It relies on the notion that very light bosons could be detected through the macroscopic forces they mediate. The latter would manifest themselves through a deviation from Newton's law at short distances, exactly in the range of the experiment. Here, we search for particles that mediate a spin-dependent force, in particular axions.
An axion would mediate a CP-violating interaction between the neutron spin~$\frac{\hbar}{2}\vec{\sigma}$ and a nucleon with mass~$m_M$ at distance~$r=|\vec{r}|$~\cite{Moody1984}:
\beq
V(\vec{r}) = \hbar^2 g_s g_p \frac{\vec{\sigma}\cdot \vec{r}}{8\pi m_M r}
\left(\frac{1}{\lambda r}+\frac{1}{r^2}\right)\,e^{-r/\lambda}.
\label{axpot1}
\eeq
%Here, $g_s\,(g_p)$ denotes the scalar (pseudo-scalar) coupling.
%%
We measure the dependence of the resonance frequencies on the neutron spin. The experiment is, therefore, slightly modified: a homogeneous magnetic guide field of 100~{\textmu}T preserves the neutron spin throughout the experiment. The neutron spin polarization is analyzed by our modified detector described above. A hypothetical spin-dependent force would change the transition frequencies. This shift is obtained by reversing the direction of both the applied guide field as well as the detector field and by measuring the difference in the count rates at the two steep slopes of the three-level resonance $\ket{1}\leftrightarrow\ket{2}\leftrightarrow\ket{3}$.
We do not observe any significant frequency shift. A fit of the strength~$g_sg_p$ and range~$\lambda$  together with all other parameters leads (at 95\%~C.L.) to an upper limit on the axion interaction strength as shown in Fig.~\ref{fig:axionexclusion}. For example, at $\lambda = 20$~{\textmu}m, an attractive coupling strength $g_sg_p > 3 \times 10^{-16}$ is excluded.
This corresponds to the most stringent upper limit from a {\it direct} search for attractive and repulsive $g_sg_p$ coupling. This limit is a factor of~30 more precise than the one derived~\cite{Baessler2007} from our previous experiment with UCN~\cite{Nesvizhevsky2005,Westphal2007}.

In summary, our experiment paves the way for the use of GRS to probe new particle physics and to search for non-Newtonian gravity with high precision.  Moreover, GRS may turn out to be an ideal tool~\cite{Abele2010} for testing hypotheses on large extra dimensions at the submillimeter scale of space-time~\cite{Arkani-Hamed1999}.

\begin{acknowledgments}
We thank G.~Pignol (LPSC Grenoble) for useful discussions, D.~Seiler (TU M{\"u}nchen) for preparing the spatial resolution detectors and R.~Stadler (Univ.~of Heidelberg) for preparing the rough mirror surface. We thank T.~Brenner (ILL) and R.~Ziegler (Univ.~of Heidelberg) for technical support.
We gratefully acknowledge support from the Austrian Fonds zur F{\"o}rderung der Wissenschaftlichen Forschung (FWF) under Contracts No.~I529-N20, No.~I531-N20, and No.~I862-N20 and the doctoral programs Solids4Fun (FWF) and
 CMS (TU-Vienna). Numerical calculations were performed on the Vienna Scientific Cluster (VSC). This work was also supported from SFB-F41 ViCoM, and the German Research Foundation (DFG) as part of the Priority Programme (SPP) 1491. We also gratefully acknowledge support from the French Agence Nationale de la Recherche (ANR) under
Contract No.~ANR-2011-ISO4-007-02, Programme Blanc International -- SIMI4-Physique.
\end{acknowledgments}

%\bibliography{qB_literature20131126}

\end{document}